%% file: main.tex
\documentclass[10pt, conference]{IEEEtran}
\IEEEoverridecommandlockouts

\usepackage{cite}
\usepackage{amsmath,amssymb,amsfonts}
\usepackage{algorithmic}
\usepackage{graphicx}
\usepackage{textcomp}
\usepackage{multicol}
\usepackage{float}
\usepackage{xspace}
\usepackage{xurl}
\usepackage[most]{tcolorbox}
\usepackage{url}
\PassOptionsToPackage{hyphens}{url}\usepackage{hyperref}
\usepackage{booktabs}
\usepackage{xcolor}
\def\BibTeX{{\rm B\kern-.05em{\sc i\kern-.025em b}\kern-.08em
    T\kern-.1667em\lower.7ex\hbox{E}\kern-.125emX}}
\begin{document}

\makeatletter
\newcommand{\newlineauthors}{%
  \end{@IEEEauthorhalign}\hfill\mbox{}\par
  \mbox{}\hfill\begin{@IEEEauthorhalign}
}
\makeatother

\title{The Green Side of the Lua}

\newif\ifanonym
\anonymfalse 

\ifanonym
  \author{\IEEEauthorblockN{Anonymous Authors}}
\else
  \author{
  \IEEEauthorblockN{André Brandão}
  \IEEEauthorblockA{\textit{University of Minho, Portugal}\\
  pg54465@alunos.uminho.pt}
  \and
  \IEEEauthorblockN{Diogo Matos}
  \IEEEauthorblockA{\textit{University of Minho, Portugal}\\
  pg55934@alunos.uminho.pt}
  \and
  \IEEEauthorblockN{Miguel Guimarães}
  \IEEEauthorblockA{\textit{University of Minho, Portugal}\\
  pg55986@alunos.uminho.pt}
  \newlineauthors
  \IEEEauthorblockN{Simão Cunha}
  \IEEEauthorblockA{\textit{INESC TEC \& University of Minho, Portugal}\\
  simao.s.cunha@inesctec.pt}
  \and
  \IEEEauthorblockN{João Saraiva}
  \IEEEauthorblockA{\textit{INESC TEC \& University of Minho, Portugal}\\
  saraiva@di.uminho.pt}
  }
\fi

\newcommand{\Lua}{\textit{Lua}\xspace}
\newcommand{\LuaJIT}{\textit{LuaJIT}\xspace}
\newcommand{\C}{\textit{C}\xspace}
\newcommand{\LuaAOT}{\textit{LuaAOT}\xspace}
\newcommand{\CodeCarbon}{\textit{CodeCarbon}\xspace}
\newcommand{\Python}{\textit{Python}\xspace}

\maketitle

\begin{abstract}
The United Nations' 2030 Agenda for Sustainable Development highlights the importance of energy-efficient software to reduce the global carbon footprint. Programming languages and execution models strongly influence software energy consumption, with interpreted languages generally being less efficient than compiled ones. Lua illustrates this trade-off: despite its popularity, it is less energy-efficient than greener and faster languages such as C.

This paper presents an empirical study of Lua's runtime performance and energy efficiency across 25 official interpreter versions and just-in-time (JIT) compilers. Using a comprehensive benchmark suite, we measure execution time and energy consumption to analyze Lua's evolution, the impact of JIT compilation, and comparisons with other languages. Results show that all LuaJIT compilers significantly outperform standard Lua interpreters. The most efficient LuaJIT consumes about seven times less energy and runs seven times faster than the best Lua interpreter. Moreover, LuaJIT approaches C's efficiency, using roughly six times more energy and running about eight times slower, demonstrating the substantial benefits of JIT compilation for improving both performance and energy efficiency in interpreted languages.

\end{abstract}


\input{chapters/introduction}
\input{chapters/methodology}
\input{chapters/analysis}
\input{chapters/conclusions}

\noindent \textit{Acknowledgments:}
This work is funded by national funds through FCT - Fundação para a Ciência e a Tecnologia, I.P., under the support UID/50014/2025 (https://doi.org/10.54499/UID/50014/2025). Simão Cunha is also funded by FCT grant 2025.04596.BDANA.

\bibliographystyle{IEEEtran}
\bibliography{biblio}

\end{document}

%% file: chapters/introduction.tex
\section{Introduction}



The United Nations' 2030 Agenda for Sustainable Development
outlines a series of goals aimed at fostering a sustainable
future by reducing the global carbon footprint~\cite{article}. A key element
in enhancing the sustainability of IT systems is the development
of energy efficient software. Energy efficient software, also
called \textit{Green Software}, refers to software that is
specifically designed and optimized to minimize the consumption
of natural resources throughout its entire lifecycle~\cite{CaleroPiattini2015}.

There are several factors that can affect the energy efficiency of
green software systems. These include poor programming
practices~\cite{Cruz2017}, the use of inefficient data structures~\cite{greens2016,10.1145/2884781.2884869},
and the selection of runtime compiler optimization flags ~\cite{compopt2012},
among others. Programming languages and their execution models can
significantly influence the energy consumption of software systems, as
demonstrated by recent studies on ranking programming languages by
energy efficiency~\cite{sle17,scp21,sle24,Gordillo2024}. These studies reveal several interesting findings. For instance, imperative and compiled languages
tend to be more energy-efficient, while slower languages can sometimes
outperform faster ones in terms of energy usage. Unsurprisingly,
interpreted languages often perform poorly in
these energy efficiency rankings.

Nonetheless, interpreted languages remain highly popular among
developers, as evidenced by their strong presence in widely recognized
programming language popularity indices, such as the TIOBE
Index\footnote{\url{https://www.tiobe.com/tiobe-index/}}.
The \Lua programming language ~\cite{Ierusalimschy1996} illustrates the contrast
between popularity and energy efficiency in programming language
adoption. \Lua has consistently ranked among the top four programming
languages that experience the most significant growth in popularity\footnote{\url{https://octoverse.github.com/2022/top-programming-languages}}.
However, from the perspective of energy efficiency, \Lua ranks
twenty-third out of twenty-seven languages in the popular energy
ranking of programming languages~\cite{scp21}. Lua consumes $45.98\times$ more energy than C, which ranks first and is considered the most energy-efficient language in the study.

It is no surprise that designers and developers of such
interpreted languages are developing and applying powerful techniques to improve both the runtime and energy efficiency of the programming language. For example, compilers are developed for interpreted languages - see, for example, the Codon compiler for Python~ \cite{codon} or LuaAOT for Lua~\cite{10.1145/3475061.3475077} that followed the old idea of using partial evaluation to produce a compiler based on the existing interpreter of Lua. Likewise, \textit{Just-In-Time} (JIT) compilation techniques are being used to improve the efficiency of such interpreted languages~\cite{jit}.

This raises several important questions that motivated the empirical
study that we present in this paper, namely:

\begin{enumerate}
\item To what extent have Lua's runtime performance and energy efficiency evolved across its versions?

\item How do just-in-time compilation techniques affect the runtime and energy efficiency of Lua compared to its interpreted execution model?

\item How do Lua's current runtime performance and energy consumption compare to those of other programming languages?
\end{enumerate}

To answer these research questions, we studied the energy and runtime performance of the 17 official Lua interpreters and 8 Lua's just-in-time compiler. The performance of all interpreters was evaluated using a well-established \Lua benchmark suite, with both execution time and energy consumption systematically measured during the execution of the programs.

Our study reveals several notable findings. In particular, all Lua JIT compilers exhibit better performance, in terms of both runtime and energy consumption, when compared to standard Lua interpreters. The best-performing Lua interpreter (Lua~5.4.7) consumes $7\times$ more energy and is about $7\times$ slower than the best LuaJIT version (LuaJIT~2.0.4). Compared to the fastest and most energy-efficient C implementation reported in~\cite{scp21}, LuaJIT consumes $6\times$ more energy and exhibits a runtime $8\times$  slower. In contrast, Lua consumes $38\times$  more energy and is $55\times$  slower than C, indicating that just-in-time compilation is a key factor in improving both performance and energy efficiency in dynamic languages.

The remainder of this paper is organized as follows: Section~\ref{sec:methodology}  details the methodology, including programs, Lua implementations, testing procedures, and environment. Section~\ref{sec:analysis} presents the results, including runtime, energy consumption. It also includes a battery endurance experiment. Section~\ref{sec:threats_to_validity} discusses threats to validity, and Section~\ref{sec:conc}  includes the conclusions.


%% file: chapters/methodology.tex
\section{Methodology}
\label{sec:methodology}

In order to perform a detailed empirical study of the \Lua language, we followed the steps described below:

\begin{enumerate}
\item We defined the set of Lua interpreters and LuaJIT compilers as well as a benchmark suite to compare the energy efficiency and runtime of all releases under study;

\item We developed an infrastructure to automate the execution of all benchmark programs across each interpreter and LuaJIT compiler release. During program execution, the infrastructure relies on Intel's Running Average Power Limit (RAPL) framework~\cite{intel_rapl_energy_reporting} to estimate energy consumption of the CPU and DRAM and records the collected energy measurements in a CSV file.

\item We performed a statistical analysis of the energy consumption of all considered Lua implementations, including correlation matrices and established sustainability metrics~\cite{powerup}.



\item Finally, we performed a battery endurance study in which we measured how many times a program could be executed by different Lua implementations and by C, starting from a fully charged battery until it was completely depleted.

\end{enumerate}

These steps are described in the next sections.

\subsection{The Lua Ecosystem}

Modern programming languages are supported by rich ecosystems consisting of (multiple) compilers/interpreters, extensive and powerful libraries, integrated development environments (IDEs), debuggers, and standardized benchmarks. Lua is no exception. In this study, we evaluate the performance of various Lua implementations by executing a set of representative Lua benchmarks. Next, we present the Lua interpreters we consider in our empirical study. After that, we discuss the benchmarks that will be used to exercise such interpreters.

\paragraph{Lua Compilers}

This study considers the most relevant Lua compilers available, including both the official reference implementation and the JIT-based LuaJIT, which is also the fastest known implementation.

\begin{itemize}
    \item \textbf{Lua (PUC-Rio)\footnote{\url{https://www.lua.org/}}:} The official reference interpreter for the Lua programming language, developed and maintained by the \textit{Pontifical Catholic University of Rio de Janeiro} (PUC-Rio). It is implemented entirely in C and is known for its simplicity and portability. Lua was first designed in 1993,  and since then, its interpreter has undergone 18 official releases~\cite{evolLua}. Therefore, analyzing the energy performance across all interpreter versions will enable us to study the evolution of Lua's energy efficiency over time and directly address our research question $RQ1$.
    
    \item \textbf{LuaJIT\footnote{\url{https://luajit.org/}}:} A just-in-time compiler for \Lua, widely recognized as the fastest Lua implementation. LuaJIT is used by organizations such as \textit{CERN}, \textit{Cloudflare}, and \textit{Neovim} due to its significant performance advantages. Therefore, including LuaJIT in our study allows us to evaluate the impact of a JIT engine on energy consumption, in contrast to a purely interpreted execution model, and thus to answer research question $RQ2$. 
    



\end{itemize}


\paragraph{Lua Benchmarks}
Throughout the years, the Lua community and academia have developed various benchmark suites to assess the performance of the language in different scenarios. Below, we briefly describe the most prominent ones:

\begin{itemize}
    \item \textbf{Lua Benchmarking:} An unofficial collection of 48 open-source benchmarks focusing on different aspects of the language, providing a broad view of Lua's performance across multiple scenarios \cite{lua_benchmarking_git}. 

    \item \textbf{Lua Benchmarks:} A more compact (and less comprehensive) set of 15 open-source benchmarks, also focusing on different facets of Lua's performance \cite{lua_benchmarks_gligneul}.

    \item \textbf{Are We Fast Yet (AWFY):} A benchmark suite compatible with 10 different languages, including Lua. It consists of 20 open-source benchmarks designed to reflect key scenarios and use cases relevant to each language \cite{10.1145/2989225.2989232}. 


    \item \textbf{LuaAOT Benchmarks:} Benchmarks presented in~\cite{10.1145/3475061.3475077}. These benchmarks are based on programs from the \textit{Computer Language Benchmarks Game} (CLBG) \cite{clbg_website}, which as also used in~\cite{scp21}.
    Some programs were excluded because they relied on non-standard libraries or made substantial use of C-based modules. This benchmark suite can be found in \cite{luaAOT}.

\end{itemize}

In our empirical study we will consider the 17 Lua interpreters, 8 LuaJIT compilers, and the LuaAOT benchmark as described next.

\subsection{Automated Energy Measurement Environment}

Once we had defined the interpreters and benchmark solutions for Lua, we tested each solution individually with each interpreter to make
sure that we could execute it without errors and that the output was the expected one.  After this preliminary analysis, we 
 select the following \Lua interpreters:
 
\begin{itemize}
    \item \textbf{The official Lua interpreter}, covering 17 releases (from \texttt{5.3.0} to \texttt{5.5.0}).
    \item \textbf{The LuaJIT compiler}: covering 7 releases (from \texttt{2.0.0} to \texttt{2.1.1}).
    \item \textbf{The LuaJIT Remake.} \footnote{https://github.com/luajit-remake/luajit-remake}
\end{itemize}

To ensure consistent and comparable execution of the Lua interpreters across all benchmarks, we adapted the benchmark suite provided by LuaAOT. Accordingly, we consider the following programs included in LuaAOT: \textit{binary-trees}, \textit{fannkuch-redux}, \textit{fasta}, \textit{k-nucleotide}, \textit{mandelbrot}, \textit{n-body}, and \textit{spectral-norm}. In addition, we use the \textit{slow} inputs described in \cite{luaAOT} to ensure that sufficient computation is performed by all Lua interpreters and JIT compilers. Furthermore, the authors provide alternative implementations of the \textit{binary-trees}, \textit{fasta}, and \textit{mandelbrot} programs for use with JIT compilers; we ensure that these implementations use the same inputs as their non-JIT counterparts. Consequently, our analysis includes seven of the fifteen benchmarks originally provided in the LuaAOT suite.

To establish a baseline for comparison with the C programming language, we utilized equivalent implementations of the benchmark programs written in C from the CLBG 23.03 version and we used the C compiler \textit{gcc 12.2.0}, as done in \cite{sle24}. These implementations serve as reference points for assessing the relative performance and energy efficiency of the Lua interpreters under study.

Having defined both the set of Lua interpreters and the corresponding benchmark programs, the next step was to collect data on the energy consumption and execution time of each benchmark under each interpreter. To measure energy consumption, we used Intel RAPL (Running Average Power Limit) \cite{intel_rapl_energy_reporting}, which is capable of providing accurate energy estimates at a very ﬁne-grained level, as it has already been proven~\cite{rapl2018}.

Each benchmark was executed 10 times. To ensure thermal consistency and prevent thermal throttling, the framework requires the processor to reach a stable baseline temperature before each run. This baseline (32.2°C) was established by measuring the idle temperature after a 60-second sleep period. A new execution begins only when the current temperature drops below this baseline plus a configurable variance ($T_{start} <= T_{baseline} + variance$). Increasing this variance allows tests to start sooner, accelerating data collection at the cost of potential thermal noise. For this study, the variance was set to 5\%.

\subsection{Experimental Equipment}

Our experiments were performed on a laptop running Ubuntu 24.04.2 LTS with the Linux 6.8.0-90-generic kernel on an x86-64 architecture. The laptop is equipped with an Intel(R) Core(TM) i7-8550U CPU @ 1.80GHz. To minimize overhead and ensure consistent performance, only the minimal necessary processes were running on the operating system during the experiments.

%% file: chapters/analysis.tex
\section{Results}
\label{sec:analysis}

After running all the CLBG problems, we analyzed the collected data to understand the performance characteristics of both the Lua interpreters and the LuaJIT compilers. This section compares the energy consumption and runtime of all Lua interpreters and LuaJIT compilers, and examines their behavior relative to C when executing the CLBG problems. The energy measurements are provided by RAPL, and we consider the Package and DRAM domains, which accurately indicate the energy consumption of the CPU and memory.

\subsection{Energy and Runtime Comparison}
Our analysis began by removing outliers using the IQR (Inter-Quartile Range) method. We then examined energy consumption trends across different versions of the official Lua interpreter and the LuaJIT compiler. The results show a clear decline in energy consumption with each successive Lua version, along with a significant efficiency advantage for LuaJIT over all other versions.


\begin{figure*}[htb!]
  \centering
  \includegraphics[width=\linewidth]{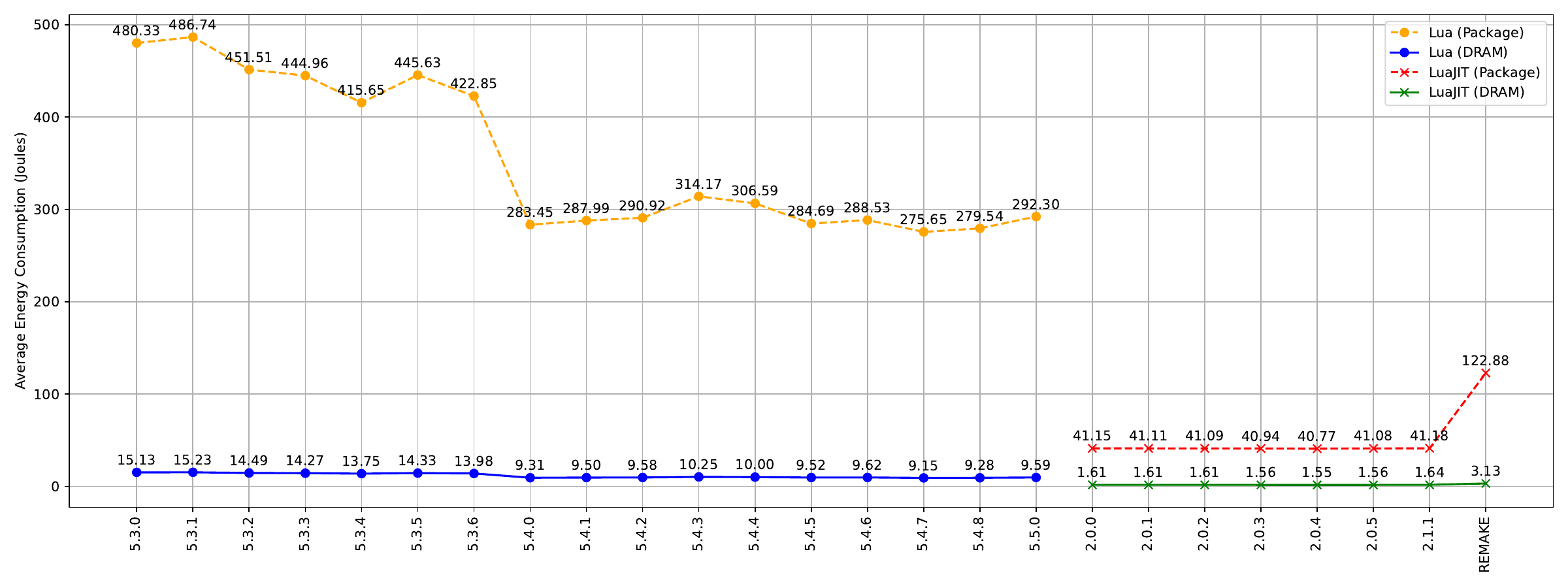}
\vspace{-1cm}
  
  \caption{Energy Consumption Across Lua Interpreters And LuaJIT Compilers.}
  \label{fig:energy_consumption}
\end{figure*}

As shown in Figure~\ref{fig:energy_consumption}, the LuaJIT version with the lowest energy consumption for both Package and DRAM domains is LuaJIT~2.0.4, with a Package consumption of $40.77$~J and a DRAM consumption of $1.55$~J. In contrast, the greenest official Lua interpreter, Lua~5.4.7, consumes $275.65$~J of Package energy and $9.15$~J of DRAM energy. When compared to Lua~5.4.7, LuaJIT~2.0.4 reduces Package energy consumption by $85\%$ and DRAM energy consumption by $83\%$.

Moreover, a historical analysis reveals substantial differences in energy consumption across official Lua releases. Notably, following the release of Lua~5.4.0, energy consumption decreased significantly for both Package and DRAM. Specifically, Package energy was reduced from $422.85$~J to $283.45$~J, and DRAM energy decreased from $13.98$~J to $9.31$~J. This represents the largest reduction observed between consecutive releases, suggesting increased attention by the Lua developers to energy efficiency considerations.

\begin{figure*}[htb!]
  \centering
  \includegraphics[width=\linewidth]{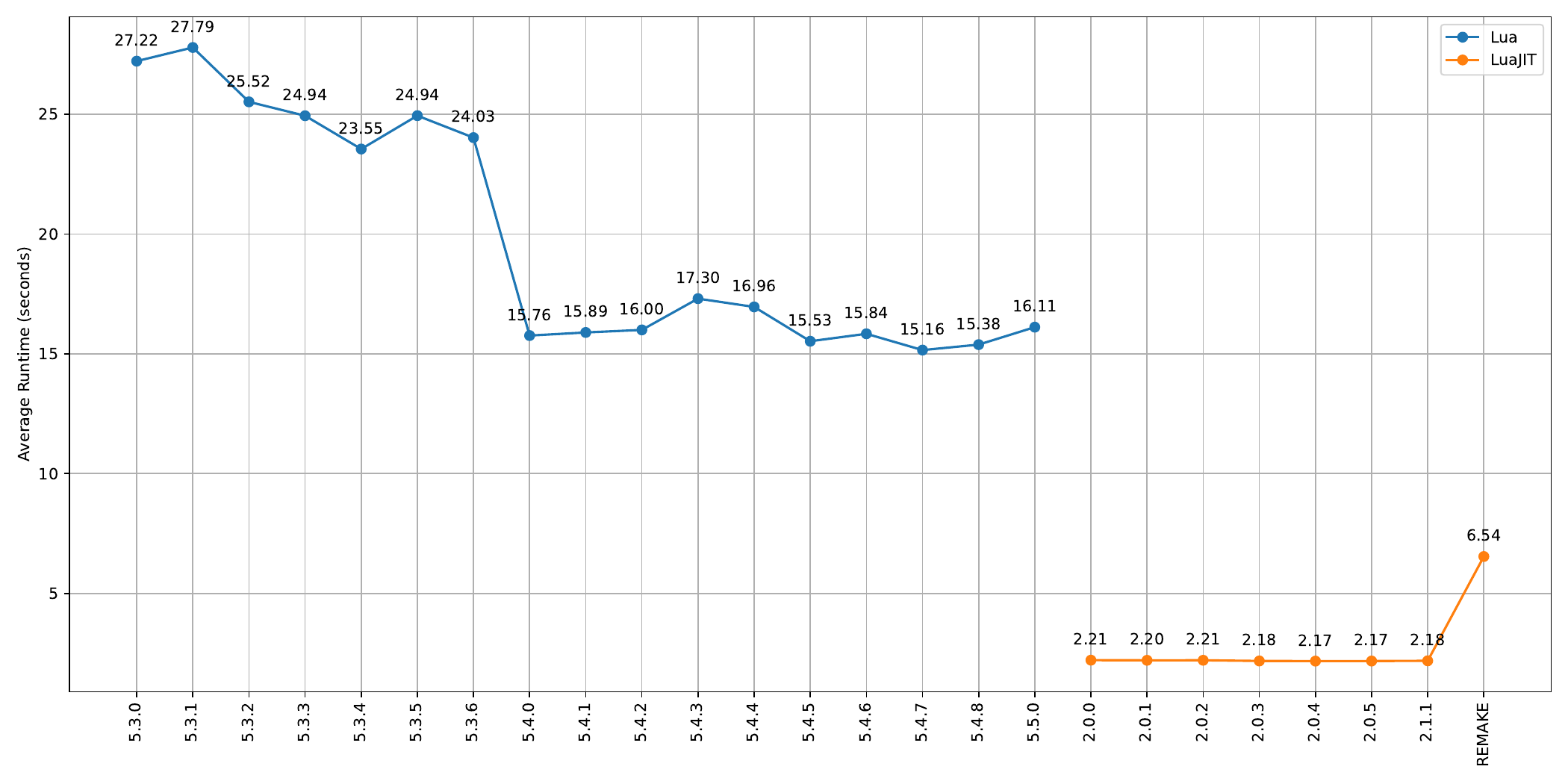}
  \vspace{-1cm}
  \caption{Runtime Across Lua Interpreters And LuaJIT Compilers.}
  \label{fig:runtime}
\end{figure*}

In addition to energy consumption, we analyzed execution time trends across different Lua versions. When comparing these two metrics, a strong alignment between energy consumption and execution time is observed. As shown in Figure~\ref{fig:runtime}, versions with shorter execution times also tend to exhibit lower energy usage, reinforcing the strong correlation between runtime and energy consumption.

As anticipated based on previous research and community feedback, the Lua community reported longer execution times with the introduction of \textit{Lua 5.4.5}~\footnote{\url{https://www.reddit.com/r/lua/comments/1dkj04j/lua_versions_546_to_515_benchmark_compared_to/}}. However, our results show that energy consumption did not increase with this version; in contrast, it improved.

The differences in energy consumption and runtime between versions \textit{5.4.4} and \textit{5.4.5} \footnote{\url{https://www.lua.org/work/diffs-lua-5.4.4-lua-5.4.5.html}, accessed December 17th 2025} could come from various causes. From our research version, Lua \textit{5.4.5} was described as a bug-fix version of the \textit{Lua 5.4} compiler, and finding the specific cause of this problem is out of the scope of this paper, still we suspect that this could happen with the usage of less complex instructions that could lead to a higher runtime and lower energy consumption.      

\subsubsection{Runtime vs Energy Consumption}

\begin{figure}[!htb]
  \centering
  \includegraphics[width=0.8\linewidth]{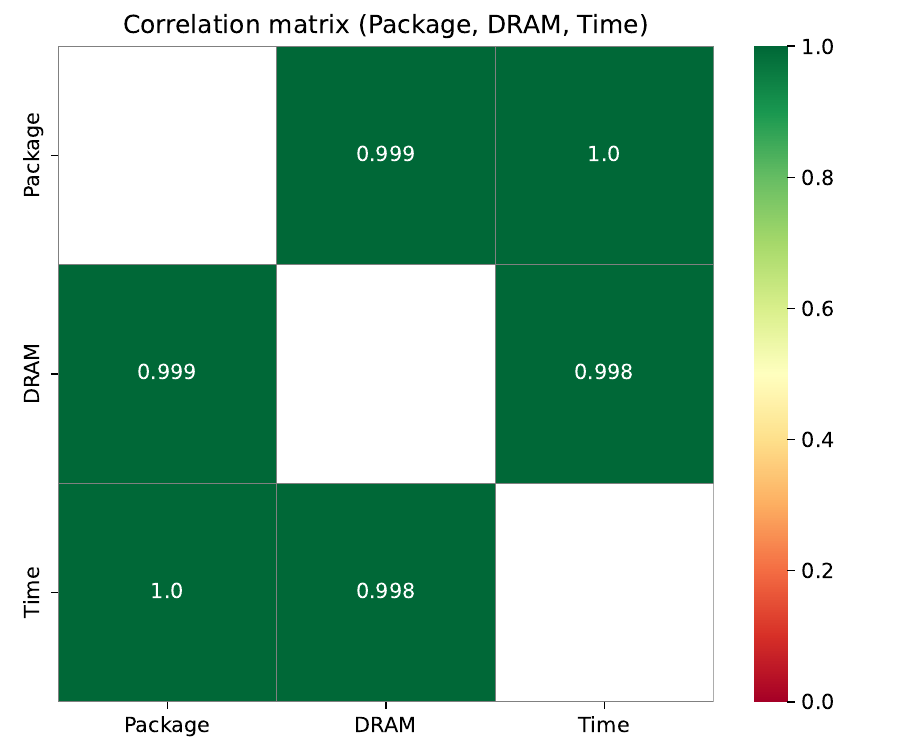}
  \caption{Correlation matrix of runtime, Package energy, and DRAM energy across all Lua versions.}
  \label{fig:energy-runtime-pairplot}
\end{figure}

As shown in Figures~\ref{fig:energy_consumption} and \ref{fig:runtime}, there is a clear relationship between execution time and energy consumption. To further investigate this relationship, considering both Package and DRAM energy consumption, we computed a correlation matrix, as illustrated in Figure~\ref{fig:energy-runtime-pairplot}. The visualization reveals a strong positive correlation among runtime, Package and DRAM energy domains across all evaluated Lua versions. This indicates that longer execution times are directly associated with higher energy consumption in both the CPU package and DRAM, suggesting that runtime is a key determinant of overall energy usage.

\subsection{Energy Consumption Across Lua Programs}

\begin{figure*}[!htb]
  \centering
  \includegraphics[width=\linewidth]{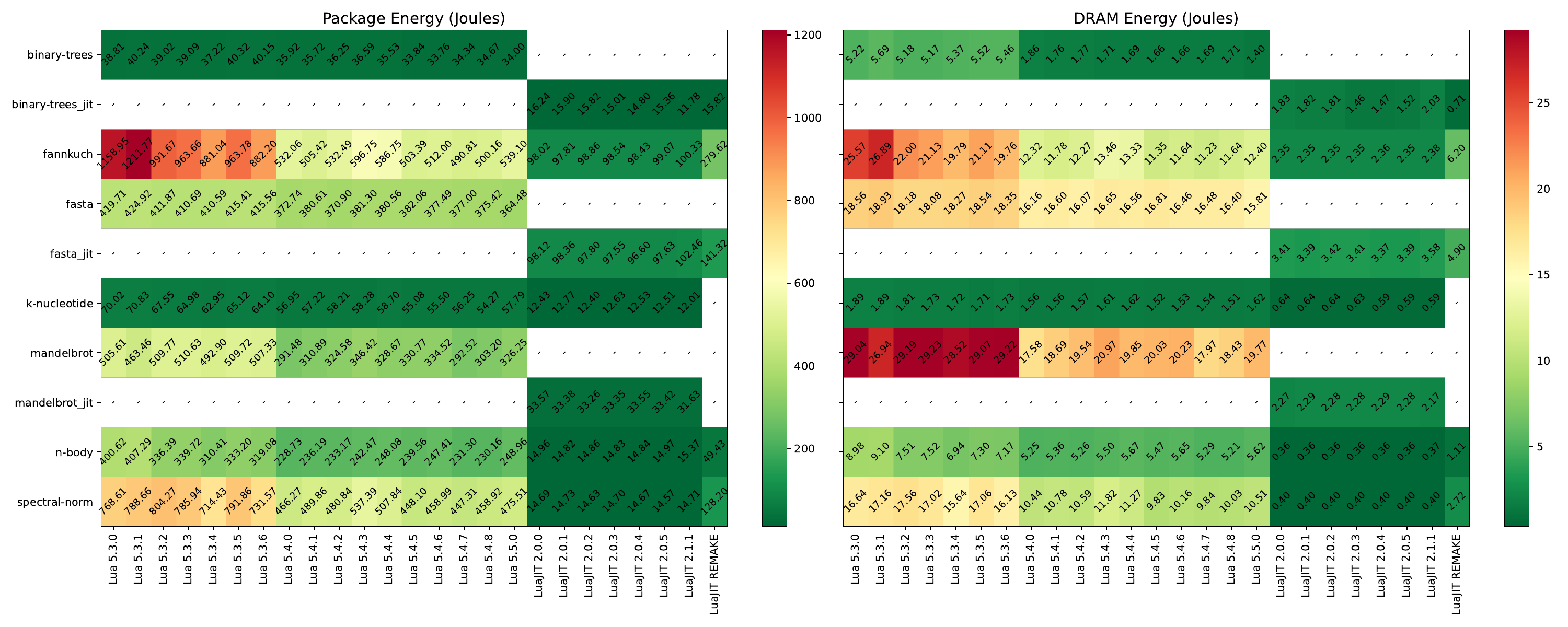}
  \caption{Energy Consumption Trends Across Lua Versions and CLBG Problems.}
  \label{fig:benchmarks}
\end{figure*}

We also independently analyzed the data for each Lua version and benchmark, as illustrated in Figure~\ref{fig:benchmarks}. Among the tests, \textit{fannkuch-redux} and \textit{spectral-norm} consistently emerged as the most demanding operations in terms of Package. These benchmarks are known to be computationally intensive across many programming languages and, therefore, do not indicate any specific disadvantage of Lua.

It is important to note that LuaJIT significantly improved the performance of the \textit{fannkuch-redux} and \textit{spectral-norm} algorithms, highlighting their strengths in highly computational workloads where just-in-time compilation can yield substantial performance gains.

\subsection{Speedup, Greenup and Powerup Metrics}

To analyze the evolution of Lua's interpreters in great detail, we used the green metrics~\cite{powerup}, namely the \textit{Speedup}, \textit{Greenup}, and \textit{Powerup} metrics. 

As detailed in Section IV of their work, we will present and explain the key equations behind each of these metrics to provide a clear understanding of how they quantify the energy-performance trade-offs observed in our experiments.

Considering two different versions of a specific software, one that is optimized and one that is not, the following metrics are defined:
\vspace{-0.3cm}

\begin{equation}
\textit{Speedup} = \frac{T_\phi}{T_o}
\label{eq:speedup}
\end{equation}

In the Speedup equation \( T_\phi \) represents the runtime of the non-optimized version, and \( T_o \) represents the runtime of the optimized version.
This metric quantifies the performance gain, where values above 1 indicate a faster optimized implementation.

\vspace{-0.3cm}
\begin{equation}
\mathit{Greenup} = \frac{E_\phi}{E_o}
\label{eq:greenup}
\end{equation}

In the Greenup equation \( E_\phi \) is the energy consumption of the non-optimized version, and \( E_o \) is the energy consumption of the optimized version.
This metric reflects energy efficiency improvements, with values above 1 suggesting reduced energy usage.

\vspace{-0.3cm}
\begin{equation}
\mathit{Powerup} = \frac{P_o}{P_\phi} = \frac{\mathit{Speedup}}{\mathit{Greenup}}
\label{eq:powerup}
\end{equation}

In the Powerup equation \( P_o \) is the power consumption of the optimized version, and \( P_\phi \) is the energy consumption of the non-optimized version.
This metric indicates how power usage changes due to optimization, with values below 1 suggesting lower average power consumption, and a value higher than 1 suggests a higher power consumption.

\subsubsection{GPS-UP Analysis}
To further contextualize the performance and energy efficiency results, we adopted the GPS-UP (Greenup, Powerup, Speedup) from \cite{powerup}.

Given the clear advantage of all LuaJIT versions (with the exception of LuaJIT-Remake) over the official Lua interpreters observed in our results, we computed the Speedup, Greenup, and Powerup metrics of the LuaJIT compiler with the lowest Package energy consumption (LuaJIT~2.0.4) relative to the remaining LuaJIT compilers and official Lua interpreters, as presented in Table~\ref{tab:jit_vs_lua}.

\vspace{-0.3cm}

\begin{table}[!htb]
\centering
\caption{Green metrics for Lua  and LuaJIT releases}
\label{tab:jit_vs_lua}
\begin{tabular}{cccc}
\toprule
\textbf{Language} & \textbf{SpeedUp} & \textbf{GreenUp} & \textbf{PowerUp} \\
\midrule
LuaJIT 2.0.4  & 1.000 & 1.000 & 1.000 \\
LuaJIT 2.0.5  & 1.001 & 1.008 & 0.994 \\
LuaJIT 2.0.3  & 1.003 & 1.004 & 0.998 \\
LuaJIT 2.1.1  & 1.006 & 1.010 & 0.996 \\
LuaJIT 2.0.1  & 1.015 & 1.008 & 1.007 \\
LuaJIT 2.0.2  & 1.016 & 1.008 & 1.008 \\
LuaJIT 2.0.0  & 1.018 & 1.009 & 1.008 \\
LuaJIT Remake & 3.010 & 3.014 & 0.999 \\
Lua 5.4.7      & 6.978 & 6.761 & 1.032 \\
Lua 5.4.8      & 7.083 & 6.857 & 1.033 \\
Lua 5.4.5      & 7.150 & 6.982 & 1.024 \\
Lua 5.4.0      & 7.260 & 6.952 & 1.044 \\
Lua 5.4.6      & 7.291 & 7.077 & 1.030 \\
Lua 5.4.1      & 7.316 & 7.063 & 1.036\\
Lua 5.4.2      & 7.365 & 7.135 & 1.032 \\
Lua 5.5.0      & 7.419 & 7.169 & 1.035 \\
Lua 5.4.4      & 7.807 & 7.520 & 1.038 \\
Lua 5.4.3      & 7.967 & 7.705 & 1.039\\
Lua 5.3.4      & 10.842 & 10.194 & 1.064 \\
Lua 5.3.6      & 11.061 & 10.371 & 1.067 \\
Lua 5.3.3      & 11.481 & 10.913 & 1.052 \\
Lua 5.3.5      & 11.483 & 10.930 & 1.051 \\
Lua 5.3.2      & 11.740 & 11.074 & 1.061 \\
Lua 5.3.0      & 12.531 & 11.781 & 1.064 \\
Lua 5.3.1      & 12.793 & 11.938 & 1.072 \\
\bottomrule
\end{tabular}
\end{table}
\vspace{-0.2cm}

\begin{figure}[htb!]
  \centering
  \includegraphics[width=\linewidth]{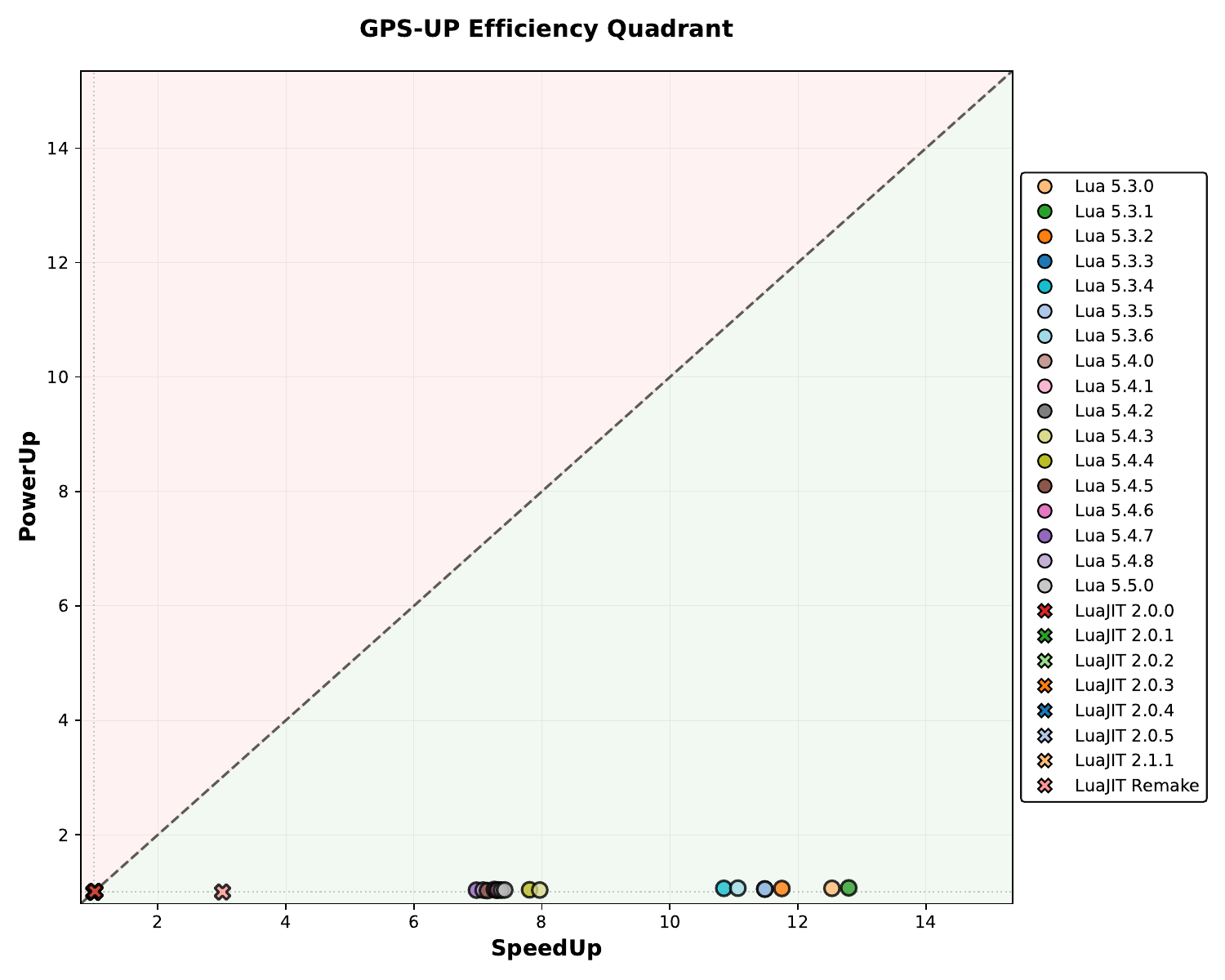}
  \caption{GPS-UP Software Energy Efficiency Quadrant: All LuaJIT versions in Zone With Less Energy and Faster Runtime Than All Lua Releases.}
  \label{fig:gps-up}
\end{figure}

Using these values, we plotted the energy–performance quadrant shown in Figure~\ref{fig:gps-up}. The results fall into Quadrant~I, indicating that all LuaJIT releases (with the exception of LuaJIT Remake) achieve both lower energy consumption and shorter execution times when compared to the official Lua interpreters. This represents the ideal scenario in energy optimization, where performance improvements are accompanied by reduced energy usage, and clearly demonstrates that LuaJIT is more energy-efficient and faster than standard Lua implementations.

Having conducted this empirical study, we are now able to address \textbf{RQ1}: \textit{How have Lua's runtime performance and energy efficiency evolved across its versions?} Based on the data presented in Figures~\ref{fig:energy_consumption} and~\ref{fig:runtime}, we observe a clear and consistent improvement in both energy consumption and execution time across successive releases of the Lua interpreter. This trend is further supported by the \textit{Greenup} and \textit{Speedup} metrics, which reflect similar progress. As shown by our results, the most energy-efficient and fastest purely interpreted version, Lua~5.4.7, achieves a Greenup of $6.761$ and a Speedup of $6.978$ relative to the fastest LuaJIT compiler evaluated, LuaJIT~2.0.4. In contrast, the earliest interpreter release considered in this study, Lua~5.3.0, exhibits a Speedup of $12.531$ and a Greenup of $11.781$ when compared to the same LuaJIT reference. These results indicate a consistent trend of simultaneous improvements in both execution time and energy efficiency across successive Lua interpreter releases.

The GPS-UP quadrant further informs our second research question, \textbf{RQ2}: \textit{How do just-in-time compilation techniques affect the runtime and energy efficiency of Lua compared to its interpreted execution model?} The results clearly demonstrate that just-in-time compilation substantially improves Lua's performance. Specifically, Lua~5.4.7 achieves a Greenup of $6.761$ and a Speedup of $6.978$ relative to the fastest LuaJIT compiler evaluated, LuaJIT~2.0.4. These findings provide a clear affirmative answer: just-in-time compilation is an effective technique for enhancing both energy efficiency and runtime performance in Lua, highlighting its considerable benefits for the sustainability of interpreted languages.

\subsection{Lua Performance Compared to C and Revisiting the Energy Ranking of Languages}

\begin{figure*}[htb!]
  \centering
  \includegraphics[width=0.7\linewidth]{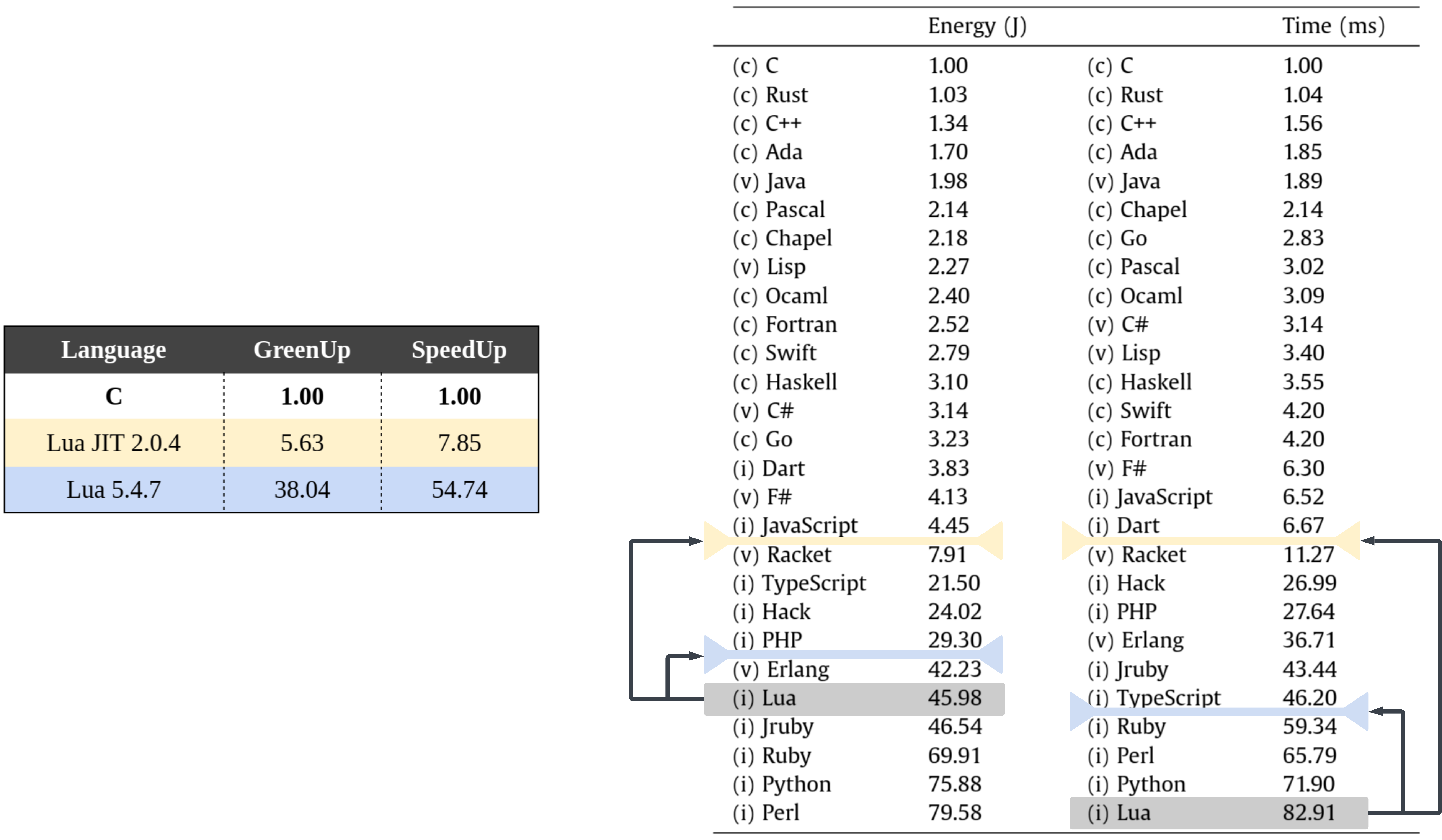}
  \caption{Lua vs LuaJIT vs C in the Ranking of Programming Languages.}
  \label{fig:lua-vs-c}
\end{figure*}

Finally, we compared our results with those obtained by running equivalent benchmarks implemented in C, i.e., the same benchmarks used for Lua but adapted to C. The comparison between LuaJIT and C, illustrated in Figure~\ref{fig:lua-vs-c}, yielded interesting insights. We observed that LuaJIT~2.0.4 consumes approximately $5.63\times$  more energy than the corresponding C implementations and is $7.85\times$  slower in terms of execution time.

Another important point of this analysis was to revisit the findings reported by the popular energy ranking of programming languages~\cite{sle17,scp21}, where Lua was poorly ranked in terms of energy efficiency and performance. The evaluation considered in such a ranking was based on version \textit{5.3.6} of Lua's interpreter, but we will compare \textit{5.4.7} as it is the fastest and greenest interpreter in our study.

To contextualize our findings, we compared our results to theirs and found the following:

\begin{itemize}
    \item In terms of energy consumption, Lua improved its ranking by 5 positions with LuaJIT (placing it between \textit{Racket} and \textit{JavaScript}), and by 2 positions with \textit{Lua 5.3.6} (between \textit{PHP} and \textit{Erlang}).
    \item In terms of execution time, Lua improved by 9 positions with LuaJIT (again between \textit{Racket} and \textit{Dart}), and by 3 positions with \textit{Lua 5.3.6} (between \textit{Ruby} and \textit{TypeScript}).
\end{itemize}

In Figure~\ref{fig:lua-vs-c}, we present the original energy rankings of programming languages from~\cite{sle17}, extended with our updated measurements for Lua. In both tables, orange highlights LuaJIT, while blue represents Lua: version \textit{5.4.7} in the left table, and version \textit{5.3.6}, as reported in~\cite{sle17}, in the right table.

For our third research question, \textbf{RQ3}, we investigate: \textit{How do Lua's current runtime performance and energy consumption compare to those of other programming languages?} In the original ranking, Lua consumes approximately $45.98\times$  more energy than C and ranks poorly alongside other interpreted languages. As shown in Figure~\ref{fig:lua-vs-c}, the use of just-in-time compilation leads to a substantial improvement. Specifically, with LuaJIT~2.0.4, energy consumption decreases to $5.63\times$ that of C, while runtime is $7.85\times$  slower. Consequently, LuaJIT rises 5 positions in the energy ranking and 9 positions in the runtime ranking. Similarly, the newer interpreter, Lua~5.4.7, climbs 2 positions in energy ranking and 3 positions in runtime ranking.

It is important to note that this comparison is not entirely reliable, as we did not use the same machine or execution environment, which naturally affects the results. Nevertheless, the outcomes we obtained still serve as a strong motivation for promoting more efficient use of the Lua programming language.

\subsection{Battery Endurance}

Estimating the energy consumption of fuel, or electrically powered machines has been practiced since the early days of the Industrial Revolution. In recent years, it became more relevant with the adoption of (fully) electric cars. In this section, we draw an analogy with the traditional method used to estimate the driving range of an electric vehicle, which consists of fully charging the battery and measuring the number of kilometers the vehicle can travel before it runs out of  battery.
We mimic this methodology by fully charging the laptop battery (the same device used in the previous study) and repeatedly running a computationally intensive program until the battery is completely depleted. The total number of executions can be interpreted as an analog of the distance traveled, measured in kilometers, in the automotive setting. 

In this study, we consider the traditional doubly recursive definition of the Fibonacci number, which is executed with the
greenest Lua pure interpreter (Lua~5.4.7), the greenest LuaJIT compiler (LuaJIT 2.0.4), and the C compiler used in the main experiment, all using the same input value (48). During the experiment, the laptop was not connected to an external power supply, and the program executed continuously in a loop until the battery was exhausted. For each execution, timestamps marking the start and end of the program were recorded in a log file.  Prior to each run, the battery was fully charged, and only the minimal necessary system processes were active. 

To reduce experimental bias, each configuration was executed three times. The aggregated results are presented in Table~\ref{tab:fibonacci_battery}. On average across the three runs, the C implementation executed the Fibonacci program 604 times, with each execution taking approximately 16.6 seconds, resulting in an average battery depletion time of 2.77 hours. 
In contrast, Lua~5.4.7 executed the program in approximately 29$\times$ more time, which resulted in about 32$\times$ fewer executions for the same energy budget. As our previous experiment showed, LuaJIT~2.0.4 is more energy efficient than Lua~5.4.7: it computed the given Fibonacci number about 5.7$\times$ more times using the same amount of energy. If we interpret the number of program executions as kilometers traveled, then, for the same energy budget, the C implementation achieved 604 executions, LuaJIT achieved 109, and the Lua interpreter only 19.

The results of our previous study are confirmed by these battery-duration tests. 
That study showed that LuaJIT~2.0.4 consumes approximately $6\times$ more energy and is about $8\times$ slower than C, 
while Lua consumes roughly $38\times$ more energy and is about $55\times$ slower than LuaJIT. 
The present tests confirm that the performance and energy-efficiency ranking among C, LuaJIT, and Lua remains consistent with the behavior observed in the main experiments.



\begin{table}[htb!]
\centering
\caption{Fibonnaci: Executions per Lua/C Language Implementation.}
\begin{tabular}{rrrc}
\hline
\textbf{Language} & \textbf{Number of Executions} & \textbf{Runtime (s)} & \textbf{Time (h)} \\
\hline
GCC 12.2.0 & 604 & 16.6 & 2.77 \\
Lua 5.4.7 & 19 & 488.2 & 2.44 \\
LuaJIT 2.0.4 & 109 & 88.2 & 2.65 \\
\hline
\end{tabular}
\label{tab:fibonacci_battery}
\end{table}

These results highlight the impact of language implementation on energy efficiency~\cite{sle17,scp21,Gordillo2024,sle24}. They also confirm that programming language implementations influence execution time, as reaffirmed in a recent work~\cite{vankempen2025itseasygreenenergy}, and therefore also impact energy consumption, given that energy can be expressed as $Energy = Power * Time$.  This is also shown in the energy ranking of programming languages~~\cite{sle17,scp21,Gordillo2024}, where faster languages are more energy efficient than slower ones. One of the few exceptions is Lua, which despite being slower than Python, has been shown to consume less energy~\cite{sle17}.



%% file: chapters/conclusions.tex
\section{Threats to Validity}
\label{sec:threats_to_validity}

We present in this subsection some threats to the validity of our study, divided into four categories~\cite{cruzes2017threats}, namely: conclusion, internal, construct, and external validity.


\paragraph{Conclusion Validity}

Although each benchmark was executed ten times without a prior analysis to determine the optimal number of repetitions, this number may be insufficient. However, the low variability observed suggests that additional runs would be unlikely to affect the conclusions. Future work could increase the number of executions to further validate these results.

\paragraph{Internal Validity} 
Measurements may be affected by background OS activity, thermal throttling, or power management changes. To mitigate these effects, we minimized background processes by disabling external services (e.g., Internet and Bluetooth), closing non-essential applications, and keeping the machine connected to power during all experiments.

{\paragraph{Construct Validity}

We used the LuaAOT benchmark suite, a curated subset of CLBG benchmarks with some programs removed due to incompatibility with the LuaAOT interpreter. This selection may not fully represent typical Lua workloads, and excluding certain benchmarks could omit cases with different performance or energy characteristics, limiting generalizability. Additionally, for the \textit{LuaJIT Remake}, the \textit{mandelbrot} and \textit{k-nucleotide} benchmarks could not be executed due to errors, and their exclusion may introduce bias in the results.

{\paragraph{External Validity}

The benchmark implementations used were those available in LuaAOT at the time of the study. As LuaAOT is an evolving project, future versions or executions on different systems may yield slightly different results. However, unless major changes occur in the language or implementations, the relative comparisons are unlikely to differ substantially. Moreover, the adopted methodology supports easy replication, as the LuaAOT benchmarks provide the necessary information to reproduce the experiments, enabling generalization and future replication by other researchers.

\section{Conclusions}
\label{sec:conc}

This paper presented an analysis and comparison of the energy and runtime efficiency of Lua interpreters and LuaJIT compilers. We considered 17 official Lua interpreter releases, the just-in-time compiler LuaJIT, and LuaJIT Remake. Each release was executed using the LuaAOT~5.4 benchmark suite, with energy consumption measured via Intel RAPL.

Our analysis revealed that consecutive releases of the official Lua interpreters exhibit decreasing energy consumption, and that there is a strong correlation between runtime and energy usage. Moreover, nearly all LuaJIT releases demonstrated substantial improvements over the interpreters. For example, the most energy-efficient and fastest Lua interpreter, Lua~5.4.7, consumes nearly $7\times$ more energy and requires approximately $7\times$ longer execution time compared to the fastest LuaJIT compiler, LuaJIT~2.0.4. We also performed a battery endurance experiment where the C implementation of the standard Fibonacci function executed the same computation 609 times, LuaJIT executed it 109 times, and the Lua~5.4.7 only 19 times all using the same battery capacity.

These results highlight the considerable advantages of just-in-time compilation techniques in enhancing the sustainability and performance of interpreted languages.

\begin{tcolorbox}[
  title=Data Availability Statement,
  colback=gray!5,
  colframe=black,
  boxrule=0.5pt,
  arc=2mm
]
The data supporting this study are openly available
in the repository \url{https://doi.org/10.6084/m9.figshare.29336132}
\end{tcolorbox}